# Deconcentration of Attention: Addressing the Complexity of Software Engineering

*by Igor Kusakov,*

*Boucherville, Canada, 2008-2012*

*http://deconcentration-of-attention.com*

*Revision: 2012-01-27*

This article attempts to describe specific mental techniques that are related to resolving very complex tasks in software engineering. This subject may be familiar to some software specialists to different extents; however, there is currently no common consensus and popular terminology for this subject area. In this article, the area is charted from a practical usability perspective.

This article also proposes to treat software engineering itself as research on human thinking because software is meant to simulate thinking.

# Contents

# Introduction

Software engineering is a complex domain. A traditional approach to a complex domain is to split it into relatively isolated parts: structurally – into components and procedurally - into life cycle phases. This way, each individual mind involved in some project can work with a reduced amount of information, which makes the work simpler. This "divide and conquer" approach has historically proven itself to be useful in disciplines such as construction and mechanical or electrical engineering.

With software engineering, however, such an approach was problematic from the start [7]. Software is too pliant, too flexible, too easy to change; thus, it is easily influenced by a constantly changing environment, such as market demands, evolving technologies, human factors, etc.

Although it might not be very obvious in small projects, developing and maintaining a medium- or large-scale project for a prolonged period of time becomes a significant challenge because there are too many moving parts which end up influencing each other despite all of the efforts to keep them isolated.

Software development is typically perceived as a process of "building" a "mechanism", which employs procedures such as "assemble" and "disassemble". However, the more complex the software becomes, the less it resembles a mechanism. A more appropriate term is "organism". An organism employs other processes aside from a mechanism, such as "growing". Thus, developing a complex software project does not resemble a process of "building a mechanism", but a process of "growing an organism". Even worse – it could resemble the process of growing an organism that is constantly mutating into something else. The "divide and conquer" approach does not work well when growing an unknown organism.

There were several major attempts to tackle the complexity of software engineering in a "traditional" manner [6]. However, none of these attempts appear to have succeeded, at least with respect to becoming popular among practitioners. Instead, we now have the "Agile Manifesto", which states "Individuals and interactions over processes and tools" [1].

We are still facing the fact that one of the biggest challenges of software engineering is the ability of an individual mind to handle extraordinarily big loads of information on a daily basis.

We do see that some people are capable of dealing with this type of challenge. There are specialists who simply "can make things work". There is no rationale behind this, just the reputation. It sometimes appears to others that such individuals could be considered "weird".

There are two reasons for studying this subject. The first reason is to be able to consciously train and hone the skill of resolving very complex and even "unresolvable" tasks.

The second reason is to avoid undesirable side effects of such activities. For example, many software specialists are affected by such things as sleeping disorders, de-socialization, emotional instability and other problems that might be directly related to the types of mental activities that they are involved in.

Approaching the subject seems to be possible via the conscious manipulation of attention. While the concentration of attention is a relatively known subject, the opposite act of deconcentration of attention appears to be the key technique for reaching specific mind resources, which are unreachable otherwise.

# 1. Background vs. Figure

Isolating figures from a background is typically considered to be the main function of attention. In some cases, there is even an attempt to reduce attention to only its isolation function. The definition of attention from wikipedia.org [17] is the following:

*"Attention is the cognitive process of paying attention to one aspect of the environment while ignoring others."*

If we start paying attention to attention itself, it is possible to notice that attention is not something abstract but rather an actual resource that has various functions that can be employed.

Consider the following visual image.

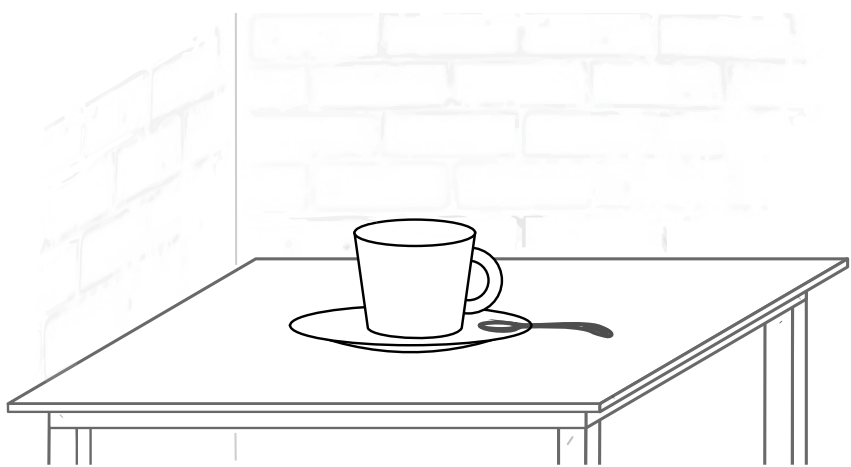

*Figure 1.1: Sample picture*

Attention, which is typically attached to where the eyes look, isolates or rather creates a figure of a cup. There are other elements in the background, which are all filtered out when the figure of a cup is formed. Attention can also create other figures – such as the figure of the teaspoon or the table.

Here is the hierarchy that attention builds:

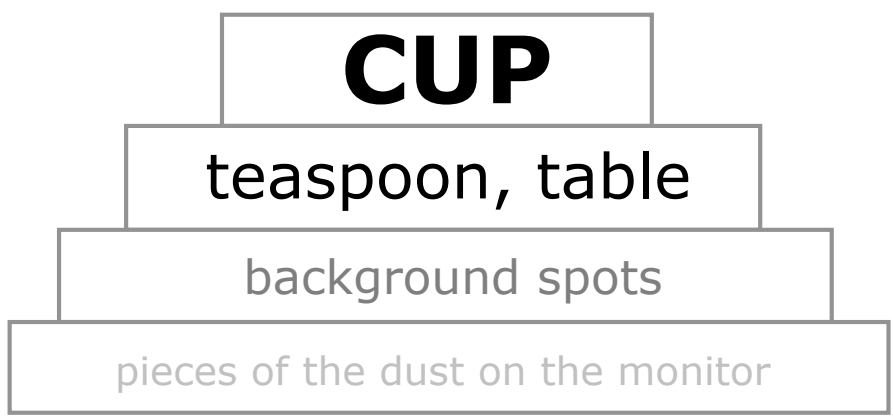

*Figure 1.2: Hierarchical perception of a visual field*

Such hierarchical perception is good for certain tasks. For other tasks, it would be preferable to perceive an image as an indivisible background, with the figure-forming function of attention stopped. Then black spots on the screen become indivisible and have the same significance as white spots and as pieces of dust on the monitor, which were initially filtered away completely. There would be no spots anymore - only one indivisible background that includes everything. This type of perception is achievable with some training.

The need for such background perception is more obvious for software engineering tasks if we see how it works in mental space.

Let's think about a "Web Server".

# Web Server

Figure 1.3: Web Server

In reality, such a thought would have a context, which turns out to be huge.

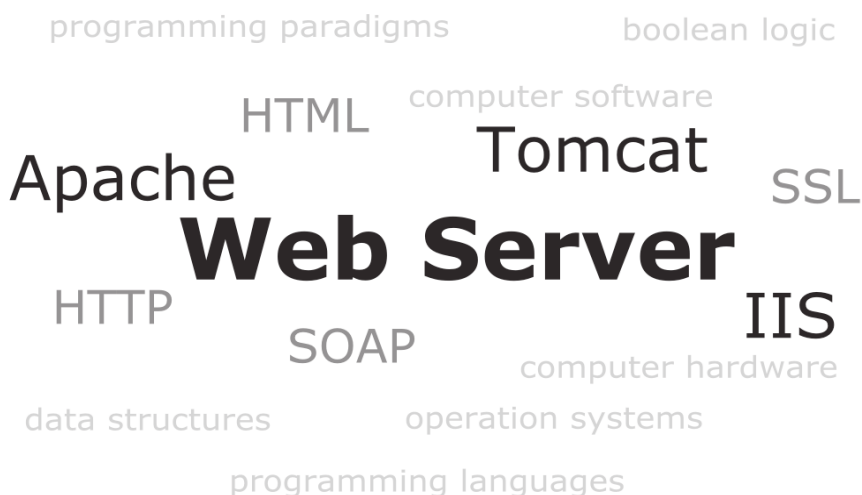

Figure 1.4: Web Server with a context

Aside from the general definition of a web server, the context includes the knowledge of some individual web servers and all of the related subjects, such as protocols, standards, programming languages, operation systems, and hardware. Overall university and school training is part of that context as well, which can also be called background.

The common phrase "someone's professional background" is a literal description of what it really is, namely a mental background.

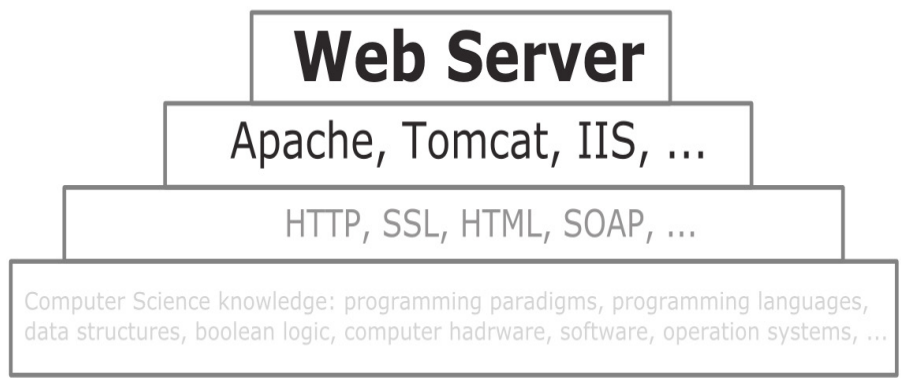

*Figure 1.5: Hierarchical perception of a "Web Server"*

In this case, "Web Server" is a mental figure isolated by attention, while all of the context information is its background. Alternatively, we can say that "Web Server" is perceived consciously, while the context remains "unconscious". Attention could jump to individual context elements if necessary. However, it is also possible to work with all of the context elements at the same time.

This special type of mental activity differs from "normal" thinking, where individual elements are serially highlighted by attention. Different thinking "modes" have different pros and cons and are applicable to different categories of tasks. The "alternative" thinking modes show themselves to be best with tasks that do not have an obvious solution or a known methodology for reaching a solution.

Let's explore how information is structured in the mental background and how it can be approached.

# 2. Pure Semantics vs. Verbalization

The process of thinking is sometimes associated with the verbal representation of information.

According to wikipedia.org [16]:

*"In popular usage mind is frequently synonymous with thought: the private conversation with ourselves that we carry on "inside our heads.""*

A more precise statement would be that verbalized text is the most common media for preserving and communicating thinking results.

Alternative media, such as symbols, pictures or even music, are not as highly evolved and typically play a complementary role to verbalized text, which is an unquestionable monopolist in representing knowledge in our days.

The monopoly of verbalization drives a tendency to associate thinking with verbalization. What also adds to this point is the constant "talking" that a human mind is trained to render all of the time within itself.

Some researchers, however, point that it is incorrect to associate thinking with verbalization [2,4]. Verbalized text can be seen as a **data format**, while the actual thinking occurs in some other form.

Let's try an exercise to explore the various alternatives a mind has.

---

**Exercise 2.1: Thinking levels**

The exercise below is performed with closed eyes.

1. Pronounce the word "red" to yourself in your mind.
2. Imagine any arbitrary red object (e.g. a hat, a car, a flag) without naming it.
3. Try to imagine an abstract red color, without naming it or imagining an associated object.
4. Try experiencing red without visualizing it, naming it or imagining associated objects.
5. Go back to step 3 – imagining an abstract red color.
6. Go back to step 2 - imagining any red object.
7. Go back to step 1 – pronouncing "red" to yourself.

Keep repeating these steps for at least a few minutes, paying special attention to step 4. The other steps are there to facilitate keeping attention focused on the selected subject (a red color).

Step 4 can be difficult in the beginning. One of the reasons for this difficulty is that the pure semantics mental sensation of the color is not something new to the mind. It has always been there but most probably was never isolated by attention before. Some struggle could be necessary to rid attention of its habit of ignoring these sensations.

One trick to help focusing attention on these sensations is to experiment with other colors and try to notice the associated mental sensations as they change. Picking other subjects (such as figures, sounds, or software engineering entities) might also help because people are different and colors may not be the easiest subject to start with for everyone.

The final goal is to be able to pick an arbitrary color as a mental sensation, without visualizing it, naming it or imagining associated objects and then to consciously either name the color, visualize it, or imagine an object.

---

The following diagram reflects the mental structure for the above exercise.

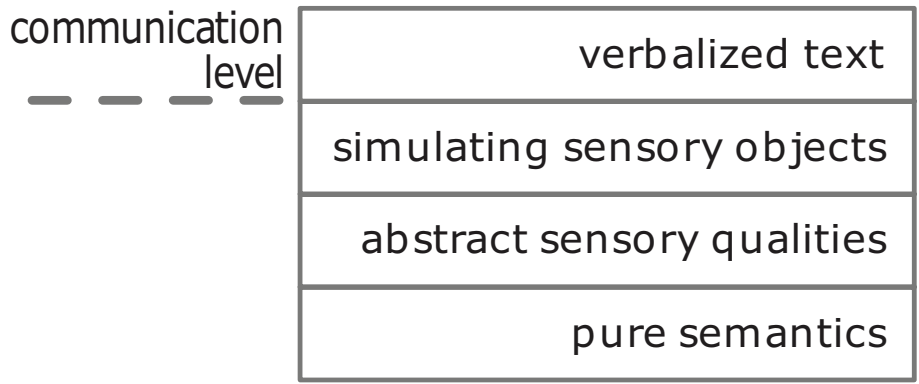

Figure 2.1: Thinking level stack

The verbalized level is what is commonly perceived as "thinking". Simulating sensory objects can be referred to as "figurative thinking".

The most interesting level for our tasks is the **pure semantics** level [15,2,4].

The distinctive feature of the pure semantics thinking level is that the information it contains is not in any way verbalized, symbolized or sensually simulated.

One common example of the pure semantics experience is the "code-switching problem", which is well known to linguists [8]. When a multilingual person must switch to a language that this person has not used for a while, the mind needs some time to "switch" to an alternative verbalization schema. There is a unique sensation of "hanging between languages". This verbalization gap is perhaps one of the best moments when many individuals had a clear sensation of the pure semantics thinking level. It is a very specific sensation when a person understands perfectly what she is trying to say but the words are not being "rendered" for some reason.

Because of its completely de-materialized nature, the pure semantics level is, therefore, the hardest level to notice consciously. The pure semantics level does produce its specific mental sensations, but they do not have direct sensory or verbal analogues.

To address this situation, a mind commonly produces two supportive phenomena: **spontaneous sensory simulation** and **spontaneous verbalization**.

The latter is the constant "talking" that a human mind typically produces within itself all the time.

Let's adapt Exercise 2.1 to software engineering. In a certain respect, reaching the pure semantics level is easier in this case because most of the software engineering subjects are immaterial and do not have any direct sensory analogs. However, it is still a good idea to split the exercise into multiple steps or phases because switching these phases makes it easier for the attention to stay focused on one subject for a long time.

> **Exercise 2.2: The pure semantics of a Web Server entity**
>
> The exercise below is performed with closed eyes.
>
> 1. Pronounce "Web Server" to yourself in your mind while doing the following three things:
>    a. Feel the vocal muscles/tongue moving as if you are really pronouncing these words.
>    b. Imagine hearing the sound of the words "Web Server" as you pronounce them.
>    c. Keep your attention on the meaning, on the sense of what a "Web Server" is.
>    Repeat the phrase like this for a several times.
> 2. Now, remove the (a) sensation that you are pronouncing these words. You still hear the imaginary sound of the words, and you still retain the meaning of "Web Server".
>    Repeat the phrase like this several times.
> 3. Now, also remove the (b) the sound of the words. What is now left is the meaning of a "Web Server" – its purely semantic mental sensation.
>    Hold this sensation for the same amount of time that you spent pronouncing the phrase, but now do not pronounce the words anymore, and do not visualize anything.
>    Experience the meaning of a "Web Server" like this several times.
> 4. Go back to step 1.
>
> Repeat the entire cycle several times, paying special attention to step 3.
>
> Once you have experimented with "Web Server" for some time, try a "Database Server". The goal is to keep these subjects in mind clearly and comfortably without naming them until there is a need. They stay in mind as meanings, senses, mental sensations, pure semantics.

The advantage of the pure semantics level is the ability to address knowledge in very efficient ways. For example, the "Web Server" entity is, verbally, simply two words. Describing its background context would require books of text. The pure semantics sensation can include the mental figure of "Web Server" along with all of its background in just one sensation, which is very impressive. Here is how it happens.

# 3. Dimensional vs. Serial

Serial thinking correlates well with the term "point of view". Dimensional or non-serial thinking represents the state of having multiple "points of view" on the same issue at the same time.

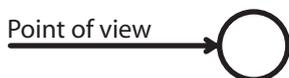 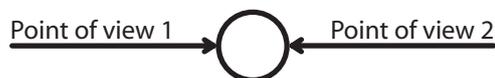

*Figure 3.1: Serial thinking*  *Figure 3.2: Dimensional thinking*

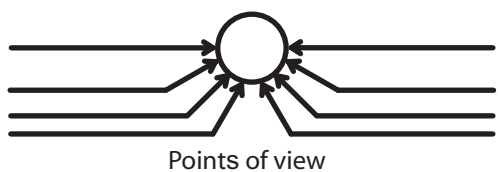

*Figure 3.3: Dimensional thinking*

The key point here is "at the same time".

Try a simple exercise:

- Say "A".
- Say "B".
- Now, attempt to say both "A" and "B" at the same time.

This exercise displays the problem that serialization causes. Saying "A" excludes saying "B", and saying "B" excludes saying "A". While it is impossible to say "A" and "B" at the same time, a mind has no such limitation and can work with both simultaneously. Thus, it appears that dimensional thinking cannot be directly represented via verbalization. However, this problem is addressed via common verbalization workarounds.

Consider the following statement:

"A and B".

This construct is a common verbalization workaround to serially represent the non-serial structure of having both A and B simultaneously.

Dimensional thinking poses potential logic problems because multiple points of view could appear to be mutually exclusive.

Consider the dialectical synthesis in philosophy [12].

Point of view 1 would be the "thesis".

Point of view 2 would be the "anti-thesis".

Having them both without logical problems would be called "synthesis".

It appears that synthesis occurs in a much simpler way at a pure semantics thinking level and, thus, in a non-serial and non-verbal form. However, verbalizing it could require significant efforts.

# 4. Background Thinking vs. Conscious Thinking

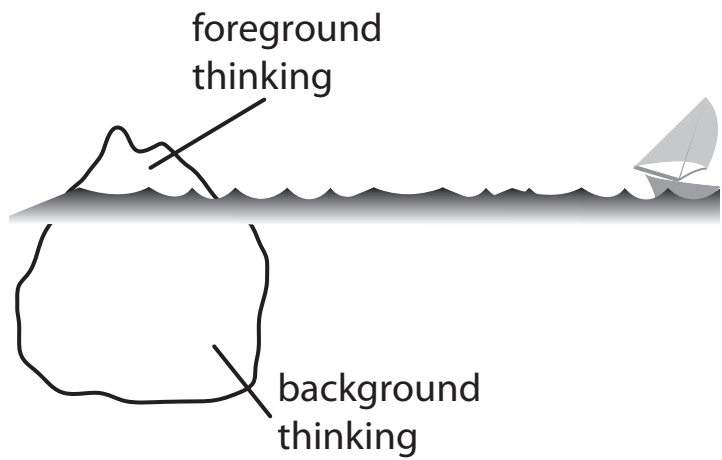

*Figure 4.1: Background thinking*

Probably one of the most powerful and mysterious abilities that a mind has is that of resolving a task without any direct involvement of conscious attention. Using this ability is also the most tiring practice, but it is quite efficient when approaching "unresolvable" tasks.

An "unresolvable" task does not necessarily mean that there is no solution. This term rather means that we do not see a solution. Practice says that, in a complex system with many alternatives, there is typically always a solution. The solution may be tricky to discover, however, because we must "scan" the hundreds or thousands of elements that are involved.

With background thinking a mind is given a goal, loaded with data, and after some time (days, weeks, months, years), the result appears. A specialist attempts to "hit" the "unresolvable" problem from various angles while the mind accumulates data. A specialist may also stop and do something else that is unrelated to the task. Although the specialist may not notice any progress at all, at some point the solution simply appears, seemingly out of nowhere.

Psychologically speaking background thinking is the use of "subconscious" resources for storing and processing data.

In most cases, the background thinking is triggered by emotions. While emotions can certainly be involved in other types of work, the link is the most obvious with background thinking. Emotions act as a "fuel" for the background thinking.

Many accomplished specialists have a strong emotional attachment to the project that they are working on. They have "a romance with the project". Or they perceive a project as their "pet".

Emotions can be triggered by all of the different possible types of reasons. They key point here is that emotions usually make this phenomenon occur. However, there could be alternative ways of triggering background thinking consciously.

The use of background mind resources appears to have its price. Background thinking is responsible for many other areas of personal life, and using it deliberately on a specific task could make it fail somewhere else. It seems as though the background processes are sorted by their importance to biological survival. Thus, what is less important fails first.

Some of the typical candidates for failure are sophisticated social behavioural patterns. Social life has many complex protocols of its own, which include behavioral rituals, dress codes, and knowledge of common discourses. Often, a technical "nerd" may give away all or some of this sophistication to free up some "mental space" for the background tasks. This could be acceptable as long as this is a conscious decision and the consequences are accounted for.

# 5. Deconcentration vs. Concentration

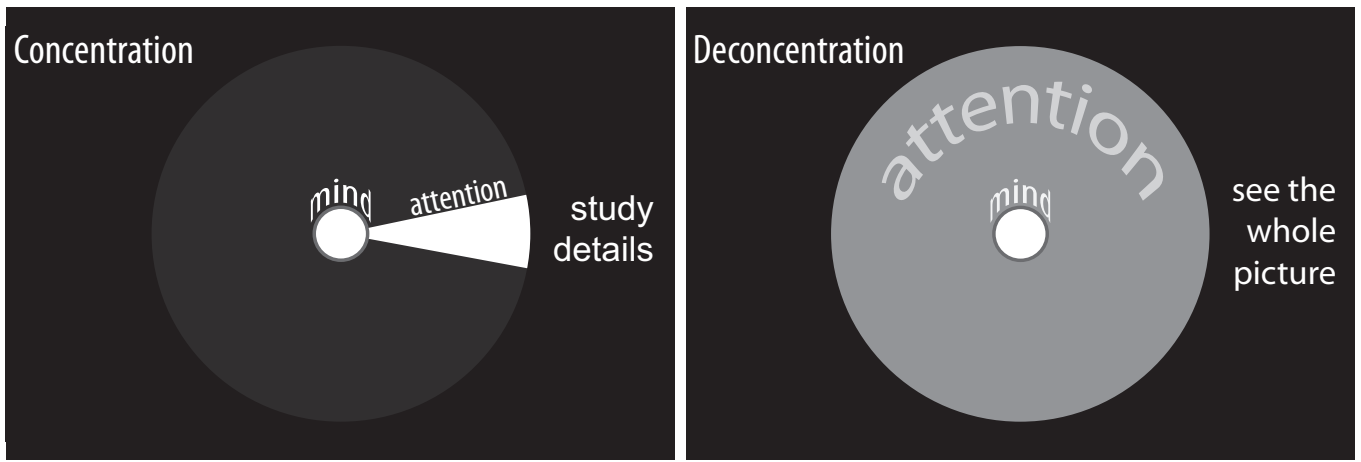

*Figure 5.1: Concentration of attention*     *Figure 5.2: Deconcentration of attention*

## 5.1 History and general description of deconcentration

*"Deconcentration of attention is opposite to concentration and can be interpreted as a process of dismantling of the figures in the field of perception and transformation of the perceptual field into a uniform (in the sense that no individual elements could be construed as a perceptual figure) background."* [3]

The term "deconcentration of attention" was introduced by Oleg Bakhtiyarov on the basis of research performed in the 1980s at the Kiev Institute of Psychology, USSR [22]. This term appears to be very well-turned and self-explanatory.

Deconcentration was originally developed *"… as part of training programs for operators in the complex, uncertain, and extreme conditions"* [22]. Back in those days, such a definition was mostly related to military and space programs. Times have changed, but in our days some software specialists might recognize "complex, uncertain, and extreme conditions" as a pretty close description of their daily work.

The concept of deconcentration states that as a mind is capable of concentrating its attention on an individual element, a mind can also deliberately de-concentrate its attention. In this way, the attention spreads equally over a certain area, which allows for an efficient approach to some tasks that would be quite difficult otherwise.

Existing studies state that concentrative attention has a maximum capacity of 5-9 objects at any moment [19]. In practice, this capacity is even less for a comfortable level. The ideal target for concentration is one single object. Maintaining concentration on multiple objects simultaneously tires the attention quickly. Addressing multiple objects simultaneously for a prolonged period of time is possible only via deconcentration.

A good example of a common activity in which a deconcentration of attention occurs to a certain extent is driving a car. Concentrating on anything for too long while driving is dangerous. Instead, a driver spreads attention on everything, without focusing on anything in particular. It is very interesting to notice how deconcentrated attention pinpoints whatever requires concentrative attention at any given moment.

It is typically considered that "automatic" or "reflectory" skills, such as driving are acquired via long repetitive training [18]. The research on deconcentration states [3] that this training time could be significantly reduced and that in some cases, a skill could even be acquired without any repetition. This quality is important for software engineering because the industry changes very fast, thus demanding serious adaptability skills.

There are two exercises in Appendix A that provide an opportunity to experience deconcentration in a visual field. These exercises could take some time to master because they expose an unusual type of conscious mental activity. The A.2 "Deconcentration over a colored number table" can provide some measurable results. The A.3 "Deconcentration over a visual field" provides a deeper experience of what deconcentration is.

While with concentration objects are sharp and clear, with deconcentration there are no objects, only the background, which could even become blurry and uncertain but being in a certain "state". This phenomenon occurs in both sensory and mental space. A pragmatic use of deconcentration is learning to properly recognize these "states" and tuning the mind to resolve specific tasks while in this perceptual mode.

One more problem with using certain deep forms of deconcentration, is that the mind stops caring or being concerned with any individual element in the field that it perceives. A concern about anything in particular immediately draws concentration toward that individual element, which breaks deconcentration. Thus, a deconcentrated mind must maintain certain indifference toward the area it is deconcentrated over. This brings up the question of how can a mind work on a task if it stops caring about it? This problem is discussed later in this article.

Deconcentration of attention in a sensory space (visual, aural or tactile) has many areas of applicability. For software engineering tasks, however, we need to be able to deconcentrate in mental space as well.

## *5.2 Deconcentration in mental space*

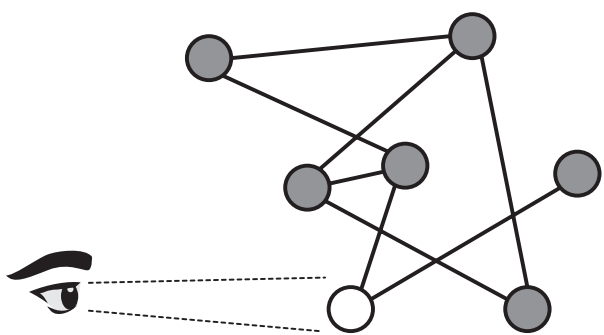

*Figure 5.3: Perceiving a complex system via concentration*

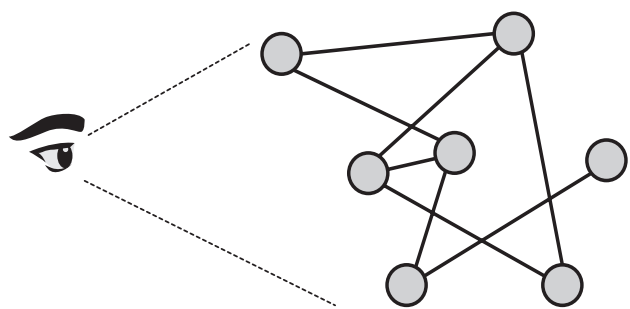

*Figure 5.4: Perceiving a complex system via deconcentration*

Deconcentration of attention in mental space appears to be the key technique to address the specific phenomena that are described in this article.

This type of deconcentration allows holding attention on a large system that consists of many elements (hundreds or even thousands). It allows the data to be structured in a dimensional manner by providing an opportunity to maintain attention on multiple points of view simultaneously. It also provides a way to interact with background thinking because deconcentrated attention is a better tool to track elusive and vague background mental signals.

One way to see the difference between the conscious and subconscious is to see it as a difference between the foreground and background. Concentration is the tool to work with the foreground. Deconcentration is the tool to work with the background.

Deconcentration in a sensory space typically starts from the perceptual boundaries and then spreads towards the center. Such direction, from the boundaries towards the center, ensures that the peripheral areas receive the same amount of attention as the center, which is the goal. Deconcentration in mental space also appears to rely on boundaries. However, while in sensory space the boundaries are defined by certain physical limitations, such as the visual field of view, the mental boundaries are quite arbitrary. The smaller boundaries are easier to handle, while wider boundaries provide more opportunities for addressing the task. It is also possible to adjust boundaries during the task resolution process, which makes things even trickier.

Examples of what could provide boundaries in software engineering are the task, the project, and the individual's overall knowledge of computer science (CS).

Deconcentration over ones total CS knowledge is required for a technical architect to produce global project decisions.

Project-based deconcentration uses existing project limitations as deconcentration boundaries. Examples of such limitations are the project programming languages, the third-party components, the protocols of exchanging data, the project architecture, and even in general "the way things are done" in a certain company. The more time that specialists spend on a project, the better is the mental image of a project, that they accumulate in their mind. At some point the quality of this mental image enables instantaneous discovery of a solution for most of the regular project tasks.

Task-based deconcentration occurs when a person does not know the project or some area well and needs to accomplish various researches that are related to the task. After a sufficient number of directions are researched, they form the deconcentration boundaries used to discover a solution.

Experiencing the actual "starting points" for these boundaries is possible via Exercise 2.2 ("The pure semantics of a Web Server entity"). Try experimenting with various subjects to become aware of their mental sensations; then they can be used similarly to the four starting points of visual deconcentration. We can start at these points and spread the attention towards the center, which is the task. There might be a problem when finding directions in mental space because it is not 2D or 3D. However, this problem arises from an attempt to visualize something in imaginary space. As long as the sensory simulation in imaginary space is avoided, there is no need to care about how many dimensions are there and how to find the directions - you just know.

Yet another approach for defining the boundaries for the mental deconcentration is to use several potentially opposite "points of view" (mentioned in chapter 3 "Dimensional vs. Serial") as the boundary points where the deconcentration starts. The following section describes what it would look like.

### 5.3 Lagrangian point

An interesting symbolic illustration of dimensional thinking and the deconcentration of attention is a gravitational field of astronomical objects, where each point of view (or idea) is represented as an object having a "gravitational" force.

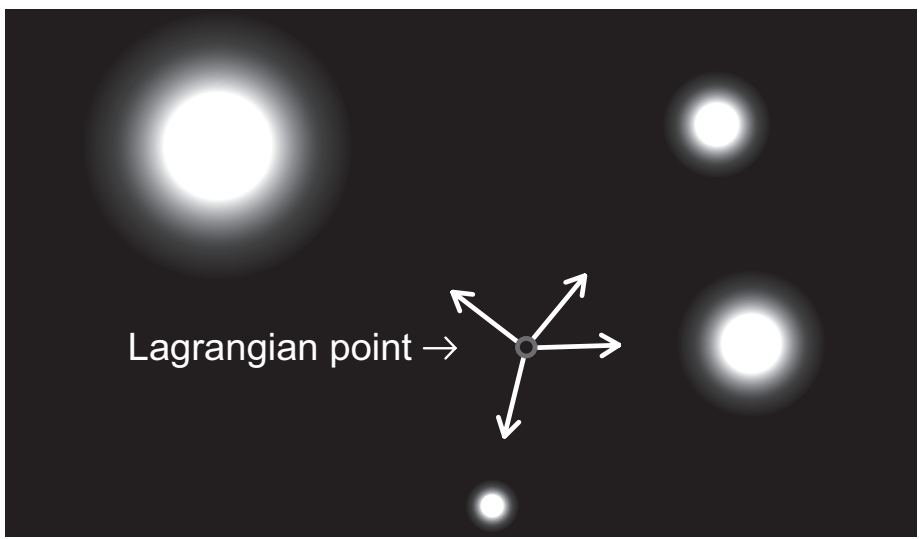

*Figure 5.5: Lagrangian point in mental space*

A solution would be the so-called Lagrangian point - a place where each idea's "gravitational" influence is balanced. It appears that the deconcentrated attention is capable of sensing this point instantaneously, without serially scanning the entire available space or all of the objects in this space.

This concept is best illustrated in a dynamic situation. When a new idea appears in mental "space" or an existing idea is proven wrong and disappears, a new "Lagrangian point" appears and can be discovered instantaneously.

When it takes time for a mind to reach this new point, this time is not about thinking but about mind adapting or becoming more comfortable with the new position. It can be an "inertia of thinking", showing the need for a more liberal approach. However, it can also be a conscious decision to approach the new position slowly and carefully in a conservative way. In any case, this concern is unrelated to the discovery of the new "Lagrangian point", which seems to occur instantaneously.

However, the verbalization of a new "Lagrangian point" is a completely different story. Verbalization could require a long description of the many ideas that are involved.

### *5.4 Examples of applying deconcentration to software engineering tasks*

One simple example of applying visual deconcentration to software engineering is a technique of finding bugs quickly in the code on a screen. Exercise A.2: "Deconcentration over a colored number table" can be directly applied to complete this task. The attention deconcentrates over the IDE editor window, where the code resides and the bugs in the code pop up instantaneously in the same way that the colored numbers pop up in exercise A.2.

An interesting side effect of such a practice is that it immediately shows the benefits of following coding standards. Deconcentrated attention recognizes both real bugs and disregardments of coding standards as some "anomalies". This makes the code that follows strict coding standards incomparably much easier to work with. In general, it appears that the more specialists rely on such "advanced" mental techniques, the more polished their code is.

While finding a bug on just one screen emphasizes the use of visual deconcentration, we typically need to work with more than that. Thus, a problem appears: how can we place all that code into mental space, where mental deconcentration can be applied? The code itself is not well suited for being stored in the mind, just as verbalized text is not the best data format in which to remember knowledge. What happens is that a specialist looks through the code and understands it. Understanding means de-serializing the code and placing it into the pure semantics mental level. While the specialist does not appear to remember the code line by line, she has a mental sensation that she knows the code.

One way to approach such a skill is to start with code that is only two screens long. Keep scrolling between the first and second screen while looking through the code and notice the sensation that you know what is on the other screen, which is not currently displayed. Then try this task with three screens and so on. Eventually this skill can be evolved to the point when you can browse very large amount of code and still somehow know what to expect from it.

At first, this knowledge "evaporates" very quickly. With repetitive work using the same parts of the code, it becomes more permanent. However, there is no need to develop a permanent knowledge of the code, unless this is really important.

There are ways to work with this "quickly evaporating" knowledge quite efficiently. The trick is to keep the code in your "now" without allowing it to go into your "past", where it attempts to become a memory. The sensation is similar to when you have started a phrase and have not finished it. Such an extended "now" can last for quite a long time, even hours, and once broken, there might not be much memory left of what was there. Among other things, such an extension of "now" requires minimizing distractions, including infrastructural/improper toolset distractions. For example, we do not typically scroll through one large file with a source code. In many cases, tracing a problem requires jumping between methods in multiple files. Thus, things such as IntelliJ IDEA's "middle click" do miracles here, allowing the method calls to be followed easily. In contrast, if a specialist needs to find those methods manually, there is a risk of forgetting why a specialist got there when she gets there.

While the related code enters into mental space, a specific goal sensation triggers the task resolution process. When sufficient data are accumulated and if the task is relatively simple, the solution appears instantaneously – first as a pure semantics sensation, and then it gets serialized into the code. In the case of harder tasks, a resolution process can be pushed into the background thinking mode.

Another interesting point is that code could be serialized directly from the pure semantics mental level, without intermediate verbalization. Such a technique produces really beautiful and efficient code. Good code is self-explanatory and does not require many verbal comments. Verbalization skills are important for communication, but in the scope of code, they are limited to commenting on non-trivial cases and contributing to the API documentation. In all other cases, suspending spontaneous verbalization and serializing the sensation of a solution directly into code appears to be the best approach. However, avoid the temptation to start an internal dialogue in a computer language, such as "thinking in java".

There is a set of specific mental sensations that are associated with all of these types of work, which are rarely named. Most of the existing terminology comes in the form of "anecdotal evidence" [11].

One good example of such "evidence" is "smelling the code", which is described by Martin Fowler in his "Refactoring" book [10]. Fowler tells a story about when he and Kent Beck worked together as consultants on various projects. At some point, they began to realize that the code in different projects has certain "smells" (mostly related to questionable coding practices). While this may seem like a funny literary metaphor, this is actually a very close description of a corresponding mental sensation, which did not have a name. Taking the closest analogy from sensory space ("smelling") appears to be a reasonable naming approach.

## 6. Will vs. Motivation

The mental techniques described above require a certain "glue" to put them together and make them work efficiently. The problem here is that motivation, as it is commonly perceived, is not sufficient for this role.

Motivation can be an initial driving force for an individual to join software engineering. Many people are attracted to this domain by all of the opportunities that it offers. For some, it could be an unmatched job market with great career opportunities and competitive salaries. We live in times when material welfare is valued. For others, the attraction could be the possibility of personal realization as a powerful hacker, a master of the digital universe.

Regardless of the initial motivation, at some point, it becomes insufficient to keep the specialist evolving. The tasks become too complex, do not seem to have solutions and are not worth the required efforts. The motivational approach makes an individual constantly re-evaluate whether the current situation is worth her efforts to reach the motivational goals. Furthermore, it is very tempting to stop because the motivational goals may have already been reached, at least to a certain extent. However, it is also possible to continue, even by squeezing the last drops from whatever motivation still exists.

At some point, the realization comes that most tasks are resolvable, regardless of how impossible they initially appeared. To achieve that resolution, a specialist may have already experienced the specific states of mind that allowed the solution to materialize. However, these very states of mind start to pose an even bigger problem. Deconcentration of attention may cause a mind to stop caring, loose common ground, and any motivation left simply vanishes.

Thus, a paradoxical situation appears when a specialist can do anything, but she does not care anymore. For some people, this scenario becomes a serious personal problem. Others attempt to adopt motivation from other areas of their lives. However, there are people who are still able to maintain their productivity, regardless of what happened with their initial motivation.

A paradoxical situation requires an equally paradoxical solution; in this case it is *will*:

"Will is a goal-oriented activity unrelated to any motivation or any external stimuli." [4]

Will is difficult to explain rationally. It is something that is experienced rather than explained. There are people who have this experience and can easily recognize it with a proper explanation. There are also people who do not have this experience. Thus, it could become an almost impossible subject to discuss.

Review the exercise "A.1: Concentration" in Appendix A. Try to keep the cube from flipping for a prolonged period of time (at least few minutes). It appears to be a difficult task because such a task requires conscious manipulation of attention, which is impossible without a certain involvement of will.

Many old cultures, including European culture up to the beginning of the 20th century, associated manifestations of will with extreme situations, such as being at war [4]. It is ironic that will is now being approached from the opposite perspective, namely by exceeding the motivational capabilities of our comparably safe and relaxed lives.

There is a temptation to try explaining will as some "high form of motivation", such as the top of Maslow's pyramid [20], which also pertains to extreme experiences and self-realization. Another example is McGregor's "Theory Y" [21], which talks about "self-motivation". The term "self-motivation" is very close to describing the manifestation of will. However, it may not be the best term because justifying "self-motivation" via a motivational paradigm poses the same conceptual problems as justifying perpetuum mobile in physics.

Motivation is essentially re-active – it is a reaction to a motive. Will is active, and it is completely detached from any forms of motivation, such as intrinsic, extrinsic, conscious, subconscious, direct, indirect or any other form of motivation.

Will is independent of both physical and psychical bodies and their manifestations. Thus, it cannot be associated with any bodily sensation or psychical function, including personal overconfidence, passion, stubbornness, or the ability to exert stress on oneself or other people. Such things are part of one's personality, and will is independent of personality.

Will appears to be related to the pure semantics mental level. Will can be perceived as an active manifestation of pure semantics. Pure semantics is experienced as a mental sensation of some "meaning" or "sense" of something. The pure semantics mental level has a unique "meaning" for everything. This "meaning" sensation turning into active force – this is will.

It appears that the impressive open-source movement that we see in our days can be explained as a manifestation of the wills of the individuals involved. While a motivational aspect does present itself here as well, it does not appear to be sufficient to justify the scope of the open-source phenomenon. An individual or a group of individuals sense the pure semantic "meaning" of a project, and this sensation is converted into the will to "materialize" that pure semantics abstraction into the world.

# 7. Software Engineering as a Research on Human Thinking

Software is meant to simulate thinking. Let's try to review the history of software engineering from this perspective.

Our culture's "official" verbal-serial-concentrative thinking has been directly mapped into a single-threaded imperative programming paradigm (with variations such as structured or conventional programming) [13].

A CPU instruction pointer acts like a concentrated attention. A CPU reads and executes a "program" or algorithm, which is basically a serial list of tasks structurally similar to a regular human language. This approach appears to be a reasonable first step to simulate thinking, as it is commonly perceived.

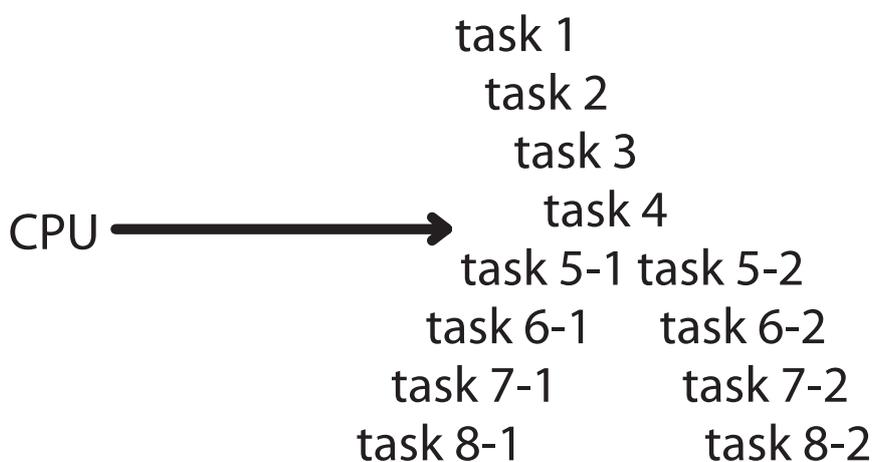

*Figure 7.1. Imperative programming*

As the complexity of software increased, the imperative programming paradigm started to show its weakness. [14]. While performing well with some mathematical abstractions, serial algorithm showed itself as a poor tool for modeling real-world domains. This looks ironic because it is our own "official" thinking mode that is under criticism.

"Dividing and conquering" the code into procedures and modules definitely helped, but the issue was not resolved conceptually because the programming paradigm still remained serial.

The dimensional approach started to shine with object-oriented programming. Despite its problems, **object-oriented programming** offered incomparably better tools for modeling non-serial structures, which ensured its dominance in the market (at least for medium- and large-scale projects).

The educational problem of the "paradigm shift" from imperative/structured programming to object-oriented programming appears to be the problem of switching from traditional serial thinking to dimensional thinking.

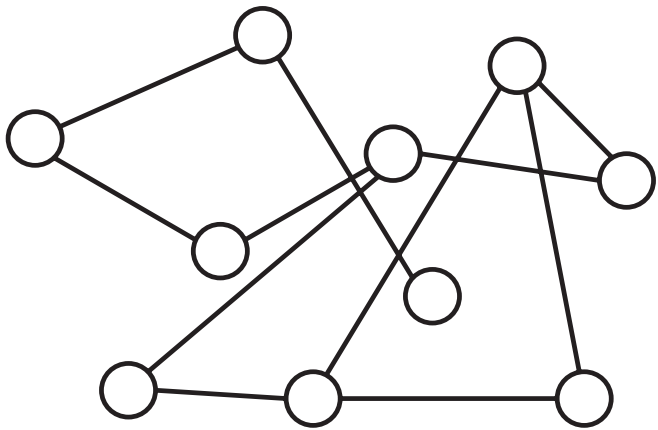

*Figure 7.2 Object-oriented programming*

According to wikipedia.org [14]:

"An object-oriented program may, thus, be viewed as a collection of interacting objects, as opposed to the conventional model, in which a program is seen as a list of tasks (subroutines) to perform."

Note how the Wikipedia author utilizes the sensory simulation terms "view" and "seen" to describe what takes place in mental space. It is useful to make a simple exercise of attempting to mentally "view" both paradigms (in any way that you prefer).

- Close your eyes and "view" conventional imperative programming. Note what you "see".
- Now, "view" object-oriented programming. Note what you "see" now.

The first one is serial. The second one is not.

One of the best books to understand what object-oriented programming is about is the "Gang of Four's" Design Patterns [9] book. The authors of the book present "patterns", which are very powerful and abstract software architectural constructs. Reaching such a level of abstraction is possible because of the quality of object-oriented programming itself. Imagine how this book would look if it was based on the imperative programming paradigm.

Verbal vs. dimensional thinking resembles different **data formats**. One format is used for interfacing, and another format is used for internal computations.

A good example of where the two data formats present themselves is the process of parsing an XML document. Such a document can be "seen" as a serial list of open/close tags with attributes and internal elements text. Alternatively, it can be "seen" as an object structure. There are precisely two APIs for parsing XML, which represent these two "visions", namely SAX and DOM. SAX is a serial approach. Anyone who has an experience of using SAX knows how cumbersome it can be. However, SAX is very efficient for certain tasks, such as the communication task of real-time processing of XML data streams of arbitrary lengths. The DOM approach is dimensional. A library code converts XML into the internal Document Object Model structure, which is not serial anymore; rather, it is a dimensional graph of objects. For many tasks, the DOM approach is incomparably easier and more efficient than SAX.

A questionable point here is that the internal memory structure of a computer is still serial, and there is still a CPU with an instruction pointer. However, this architecture can be considered a hardware limitation. The hardware still reflects an imperative programming paradigm, and radical paradigm shifts in hardware are yet to come, maybe with quantum computers.

Computers and human minds appear to be facing a similar conceptual conflict between hardware and software but in the opposite ways. Computer hardware is designed to work in a verbal-serial mode from the start. However, the software pushes it into non-serial tasks. Human mind "hardware" appears to be originally dimensional and non-verbal, but it is pushed into working in a verbal-serial mode.

## Conclusions

The scope of this article is not sufficient to thoroughly cover its subject. This is probably acceptable because the subject currently needs to be surfaced, not covered.

The assumption is that the phenomena described are natural and do manifest themselves, although they are rarely approached consciously. Thus, this article attempts to provide a list of pointers: what to look for and where to look.

The most natural approach to explore this area is to attempt to track and recognize these phenomena in our daily work. Such conscious recognition has an ongoing evolving effect. These phenomena will slowly start to convert from being some "runaway kids", living in the shadows never lit by consciousness, into "rightful citizens" of the psychical space, with overall balancing and efficiency-improving effects.

With more enthusiasm, special exercises can be practiced. Several exercises are presented in this article and they could, of course, be modified, extended and fine-tuned for specific needs. Other than that, there are already some studies and groups of exercises available on improving the efficiency of mind. I personally prefer how Oleg Bakhtiyarov [2,3,4] structured this area from the perspective of technology or, as he calls it, psychotechnology. This article can be seen as a popularization and adaptation of Bakhtiyarov's works to software engineering needs.

Bakhtiyarov himself is targeting much broader goals and is developing technologies that could be applied in almost any area [22]. One of his goals is to be able to train a "universal specialist" - a specialist that could efficiently adapt to resolve any category of tasks. While such goals are very impressive, I am interested in bringing these concepts and practices into a relatively self-contained but pragmatic and problematic area such as software engineering. This may benefit both the software engineering domain by addressing its needs and the research itself because this research would have an excellent testing ground.

Many of Bakhtiyarov's practices are structured around working with perceptual fields, such as visual, aural, and tactile fields, and around how this perception interacts with the pure semantics level. This is very good for understanding the subject in general, and it has a big potential for applicability in many areas. However, software engineering subjects are mostly immaterial and do not have direct sensory analogs. Thus, we need more practices and technologies to work with the mental field specifically, unrelated to any regular perception.

Information technology (including software engineering) challenges most of the traditional practices and approaches that it attempts to adapt. Traditional practices are not working as expected or not working at all. This appears to be reasonable because information is a matter quite distinct from what traditional disciplines are usually dealing with.

What is even more intriguing is that information technology appears to be challenging the traditional ways in which our mind behaves and the way we think that we think. This is a very interesting subject to explore.

## Acknowledgements

I would like to thank my colleagues throughout my career. It is witnessing of how software engineers work and informal discussions that lead me to the most of the conclusions reflected in this article. I especially thank for those doomed and unrealistic projects we worked at that helped uncovering what individuals are really capable of.

I would like to thank Oleg Georgyevich Bakhtiyarov for his ongoing research, for reviewing this article, and for supporting its creation - this was very important to me. I would like to thank Dmitry Yakshin for the help with graphic design and document formatting. And I would like to thank Manuel Waelti for reviewing this article, providing many useful comments from the perspective of psychology and linguistics, and providing moral support throughout this article creation process.

## Copyright Notice

# Appendix A: Exercises

Here are three examples of general exercises for practicing concentration and deconcentration. These exercises are adapted from those originally developed by Oleg Bakhtiyarov [2,3,4] for his training programs.

Both concentration and deconcentration skills must be practiced together to avoid undesirable side effects. These exercises are applied in the visual sensory field, which is probably the simplest area to start from. After the unique sensation of each "mode" of attention is experienced, it is easier to recognize it in mental space.

## *A.1 Concentration*

Here is a 2D image of a 3D cube.

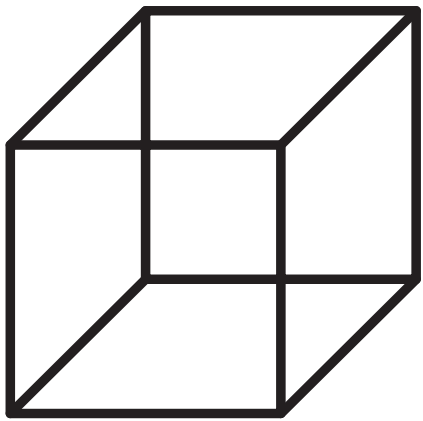

*Figure A1.1: A 2D image of a 3D cube*

There are two ways of perceiving this image as a 3D cube:

1) With the bottom-left side of the cube in front.
2) With the top-right side of the cube in front.

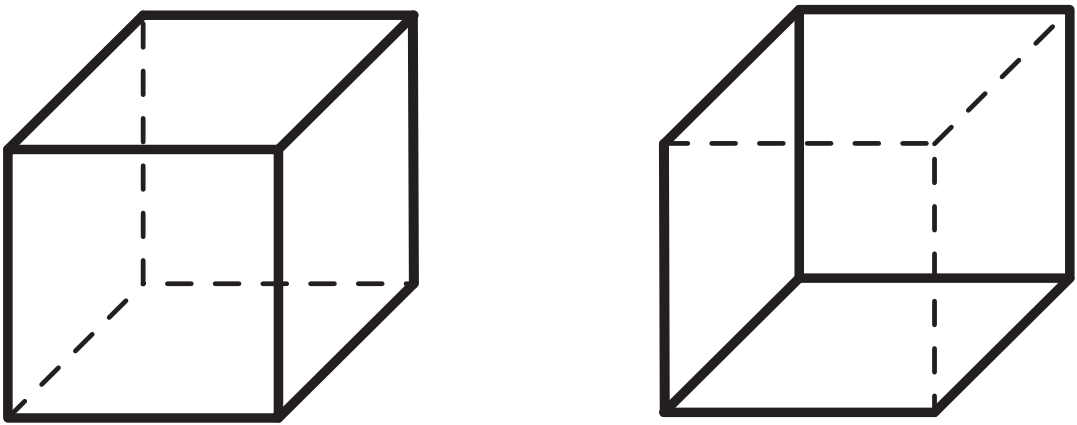

*Figure A1.2: Two ways to see the cube*

Select one way to see the cube in Figure A1.1 and maintain that choice for a prolonged period of time (for at least a few minutes).

After a short time, you may notice that the cube starts "flipping" spontaneously. Another side of the cube comes to the front without you allowing that to happen.

The exercise is to avoid this spontaneous "flipping" for as long as possible.

You may notice that commanding perception in this way is not a simple matter. This exercise demonstrates how will is involved in such activities.

If you master the ability of keeping one cube from flipping, then try this exercise with two cubes side-by-side. Hold cubes in the opposite states, for example the left one with the bottom-left side in front and right one with the top-right side in front.

With two cubes, you can also shift your eyes focus until you see 3 cubes, with the central cube being very realistic. This technique deepens the concentration on the cube and provides a clear criterion regarding whether you still keep your concentration or not. As soon as you start loosing the perception of the central cube, it means you are loosing your concentration.

The most interesting and difficult variation of this exercise is to be able to see a cube that is flipped in both ways at the same time. If you manage to do so, the experience is very unusual. You will basically see an object that is not 2D or 3D and cannot exist in the regular world.

### A.2 Deconcentration over a colored number table

A good exercise to experience specific deconcentration sensations and to see how deconcentration is used to resolve a task is to use a colored number table. This exercise also allows measuring the speed improvement between resolving the task via the means of concentration and via deconcentration.

| 16 | 13 | 9 | 23 | 1 | 5 | 9 |
|---|---|---|---|---|---|---|
| 22 | 4 | 6 | 13 | 19 | 22 | 11 |
| 3 | 15 | 2 | 12 | 24 | 5 | 18 |
| 17 | 12 | 18 | 8 | 15 | 20 | 17 |
| 25 | 3 | 20 | 1 | 4 | 23 | 6 |
| 7 | 10 | 8 | 16 | 7 | 21 | 14 |

Figure A2.1: Colored number table

The task is to find black numbers from 1 to 25 and red numbers from 24 to 1: 1 black, 24 red, 2 black, 23 red, and so on.

If you want to measure the difference in speed then try accomplishing this task by searching for numbers in a regular manner using the concentration of attention. Measure the time that it takes to complete the task.

Now, try to deconcentrate over the table area. Because this is probably a new type of conscious mental activity, this task may require some effort to achieve at first. It becomes easier with practice.

- Look at the area in the center of the table without focusing on anything in particular. The eyes may even become unfocused, looking nowhere but in the general direction of the table.
- While looking this way, focus the attention on the top-left corner of the table. The important point here is to separate the attention from where the eyes look. The eyes should keep looking in the direction of the table center in a relaxed and unfocused way during the whole exercise.
- While keeping your attention on the top-left corner of the table, add the top-right corner to your attention without moving your eyes. Now, you have two areas that are highlighted with attention. Remember not to move your eyes to look there. Try not to move your eyes at all – relax them.
- Add the bottom-left and bottom-right corners to your attention. Now the attention highlights 4 corners of the table.
- Spread your attention over the outermost row of the table. Now, the attention forms a frame of a square.
- Start covering the entire table area with attention, starting from the outside frame and going toward the center. When the entire table area is covered with attention, maintain this state for some time. The eyes are still not focused on anything and are looking somewhere in the direction of the center of the table. At this time, the table may look blurry or even visually disappear, and various visual effects may occur.
- Next, try searching for a number. Do not move your eyes to look for it. Just make yourself very interested and very concerned with finding, say, 24 red. Deconcentrated attention pinpoints the number instantaneously – it pops up immediately from the table.
- When the number pops up, do not move your eyes to see it; just notice it with your peripheral vision. Then find the next number in the same way.

The "number popping-up" visual effect is an impressive phenomenon that is relatively easy to experience. You can now measure the time that it takes to find numbers via deconcentration.

One issue you may notice, is how the attitude of the mind changes. While being deconcentrated, the mind is indifferent towards the table and its contents. To find a number, you need to become concerned with that specific number. Try making yourself believe that the number is important and that you really need to find this number now. As soon as you start really caring about the number, it gets pinpointed instantaneously. However, this action threatens to break your deconcentration. Thus, to retain your deconcentration, you need to become indifferent again. Then, you need to become concerned again with finding another number.

Deconcentration exercises are not learned solely via automatic repetition. They require active personal involvement and personal interest throughout the whole process. [3].

### *A.3 Deconcentration over a visual field*

Practicing deconcentration is possible and even more natural using the entire visual field of view.

- The eyes look in front without focusing on anything. The eyes stay this way throughout the entire exercise.
- Attention locates the left-most object in the peripheral field of view. If there is no object there, then the attention covers only the left-most spot. There are ways to trick the attention into becoming interested in this particular point. For example, boys can imagine a potential threat from that point, such as an alien invasion. Girls may start expecting a chocolate bar to jump out of that spot.
- The right-most peripheral spot is added in the same way. Now, attention is tracking two areas, which is already a form of deconcentration.
- Then the top-most and bottom-most spots are added so that the attention is tracking four points.
- Then the entire peripheral borderline becomes covered with attention. The attention now forms an ellipse.
- Then, the attention spreads over the entire field of view, moving from the edges to the center.

Correctly performing this exercise can lead to a group of specific phenomena. Ideally, the attention stops creating objects, and the visual field is perceived as one indivisible background. Various visual effects may occur, such as seeing the world as a chaotic set of colors. The visual field could also become covered in white or gray fog. These phenomena are normal because we should not perceive any objects; rather we perceive the overall "state" of the visual field.

The idea is that a mind is capable of addressing certain tasks efficiently with this type of perception. For tasks that require working with multiple objects at the same time or tracking barely noticeable signals, deconcentration of attention is the proper tool.